\documentclass[twocolumn,showpacs,prl]{revtex4}
\usepackage{graphicx}
\usepackage{dcolumn}
\usepackage{bm}
\begin{document}

\title{The Theoretical Power Law Exponent for Electron and Positron Cosmic Rays: \\ 
A Comment on the Recent Letter of the AMS Collaboration}
\author{A. Widom and J. Swain}
\affiliation{Physics Department, Northeastern University, Boston MA USA}
\author{Y.N. Srivastava}
\affiliation{Physics Department, University of Perugia, Perugia IT}

\begin{abstract}
In a recent letter, the AMS collaboration reported the detailed and 
extensive data concerning the distribution in energy of electron and 
positron cosmic rays. A central result of the experimental work resides 
in the energy regime $30\ {\rm GeV} < E < 1\ {\rm TeV}$ wherein the 
power law exponent of the energy distribution is measured to be 
$\alpha ({\rm experiment})=3.17$. In virtue of the Fermi statistics 
obeyed by electrons and positrons, a theoretical value was predicted 
as   $\alpha ({\rm theory})=3.151374$ in very good agreement with 
experimental data. The consequences of this agreement between 
theory and experiment concerning the sources of cosmic ray electrons 
and positrons are briefly explored.
\end{abstract}


\maketitle

\section{Introduction \label{intro}}

In a recent letter\cite{Ting:2014}, the AMS collaboration reported the energy 
distribution of the lepton sector, i.e. the energy probability density 
\begin{math}  \rho (E) \end{math} of the electron and positron 
contribution to cosmic rays. In particular, the AMS collaboration measured the 
slope of the log-log plot of the energy distribution 
\begin{math}  \rho (E) \end{math} versus the single cosmic ray particle energy 
\begin{math}  E \end{math},
\begin{equation}
-\alpha (E)=\frac{d \ln \rho (E)}{d \ln E}=
\left[\frac{E}{\rho (E)}\right]\left[\frac{d \rho (E)}{d E}\right] .
\label{intro1}
\end{equation}

We have previously argued\cite{Widom:2014} that for cosmic rays 
emitted as a particle wind  evaporating from a compact stellar source, 
such as a neutron star, the function 
\begin{math}  \alpha_{\rm Fermi} (E)  \end{math} may be 
computed from quantum statistical thermodynamics. The argument is 
reviewed in what follows below. 
For fermions -in the asymptotic high energy region- we predicted 
that the power law exponent would be given by 
\begin{equation}
\alpha_{Fermi} ({\rm theory})=3.151374,
\label{intro2}
\end{equation}
while in the experimental region, 
\begin{math} 30\ {\rm GeV} < E < 1\ {\rm TeV}  \end{math}, we have 
\begin{eqnarray}
\alpha ({\rm experiment})=3.170\  \pm 0.008 {\rm (stat+syst)}\nonumber\\
  \ \ \ \ \ \ \pm 0.008 {\rm (energy\ scale)}, 
\label{intro3}
\end{eqnarray}
wherein the agreement between Eqs.(\ref{intro2}) and (\ref{intro3}) is more 
than satisfactory. Incidentally, were leptons a classical relativistic Maxwell-Boltzmann
gas, the index would have been
\begin{equation}
\label{intro4}
\alpha_{MB} = 3,
\end{equation} 
instead of the corresponding Fermi-Dirac index given in Eq.(\ref{intro2}). The difference
is small and positive
\begin{equation}
\alpha_{FD} - \alpha_{MB}\approx\ + 0.151374,
\label{intro5}
\end{equation} 
that properly accounts for the repulsion between fermions inherent in their quantum statistics.

In the concluding section, we briefly discuss the implications 
of this agreement between theory and experiment. 

\section{Statistical Thermodynamics \label{st}}

The energy distribution of particles evaporated from compact stellar objects is 
determined by the entropy per evaporated particle  
\begin{math}  s(E) \end{math} 
via \begin{math}  \rho(E) \propto \exp [ -s(E)/k_B ] \end{math}. Together  
with Eq.(\ref{intro1}), we thereby have    
\begin{equation}
\alpha (E)=\left[\frac{E}{k_B}\right]\frac{ds(E)}{dE}
 = \frac{E}{k_BT(E)}\ ,
\label{st1}
\end{equation}
wherein the thermodynamic relationship 
\begin{math} T=dE/ds  \end{math} has been employed. It is computationally 
more simple to find the energy as a function of temperature 
\begin{math} E(T) \end{math} and later find the temperature as a function 
of energy \begin{math} T(E) \end{math} as an inverse function. In virtue  
of a relativistic ideal Fermi gas with a density of states per unit energy per 
unit volume \begin{math} g(\epsilon)  \end{math}, one obtains  
\begin{eqnarray}
g(\epsilon )=\frac{\epsilon \sqrt{\epsilon^2-m^2c^4}}{\pi^2\hbar^3c^3} 
\ \ \ {\rm for} \ \ \ \epsilon \ge mc^2 ,
\nonumber \\ 
f(\epsilon) = \frac{1}{\exp (\epsilon /k_BT)+1}\ ,
\nonumber \\ 
E=\overline{\epsilon}
=\frac{\int \epsilon g(\epsilon)f(\epsilon)d\epsilon}
{\int g(\epsilon)f(\epsilon)d\epsilon} \ .
\label{st2}
\end{eqnarray}
For the ultra relativistic regime wherein  electron mass effects 
are small, i.e.  
\begin{math} mc^2 \ll k_BT    \end{math} ,  
\begin{math} mc^2 \ll E  \end{math} and 
\begin{math} (E/k_BT)={\rm constant} \end{math}, our theoretical 
prediction in Eq.(\ref{intro2}) is recovered from 
Eqs.(\ref{st1}) and (\ref{st2}). 

\section{Radiation Damping \label{rd}}

If the energy distribution of evaporating electrons and positrons from 
compact stellar sources is as described above, then the question arises 
as to whether this energy distribution changes appreciably due to radiation 
damping as these cosmic leptons propagate from the source to the laboratory 
detectors built within our solar system. We here note that the accelerations 
of these charged particles due to random cosmic electromagnetic fields of the 
order of micro-Gauss do not radiate appreciable energy for propagation 
distances of galactic proportions. 

To see what is involved, consider an electron or positron with 
energy \begin{math} E=mc^2 \gamma \end{math} moving along a circular arc 
in a magnetic field \begin{math} {\bf B} \end{math}. Due to radiation damping,  
it is {\em well known} that the radiation energy loss obeys\cite{Landau:1975} 
\begin{equation}
\frac{d\gamma }{dt}=-\frac{1}{\tau}(\gamma^2-1),
\label{rd1}
\end{equation}
with a characteristic time scale \begin{math} \tau  \end{math} determined by 
\begin{eqnarray}
\frac{1}{\tau}=\frac{2}{3}\left(\frac{e^2}{mc^3}\right)
\left(\frac{eB}{mc}\right)^2 ,
\nonumber \\ 
\tau  \approx 5.15868\times 10^{20}\ {\rm sec}
\left(\frac{\rm 10^{-6}\ Gauss}{B}\right)^2 .
\label{rd2}
\end{eqnarray}
With an initial energy \begin{math} E_i=mc^2\gamma_i=mc^2\coth \chi  \end{math},
the exact solution of the radiation energy loss is thereby 
\begin{equation}
\gamma (t)=\coth\left(\frac{t}{\tau}+\chi\right)=
\left[\frac{\tanh(t/\tau )+\gamma_i}{1+\gamma_i \tanh(t/\tau )}\right].
\label{rd3}
\end{equation}
It is thereby evident that 
\begin{equation}
1\gg \gamma_i \tanh(t/\tau)\ \ {\rm and} \ \ \gamma_i\gg 1 
\ \ {\rm implies} \ \ \gamma(t)\approx \gamma_i.  
\label{rd4}
\end{equation}
For the energy range of importance in the AMS experiment and for cosmic magnetic 
fields of the order of micro-Gauss, the time of flight for an electron or positron 
emitted from a compact source to a detector within our solar system without 
appreciable radiation damping of energy is the upper time limit 
\begin{math} t < t^* \sim 10^{14}\ {\rm sec} \end{math}. The time 
\begin{math} t^* \end{math} is by a large margin more than the time taken for a 
speed of light signal to transverse our galaxy.

\section{Conclusion \label{conc}}

In the baryon sector, the heavy nuclear cosmic ray particles can 
be bosons or fermions, the least massive cosmic ray particles being protons that are 
fermions. Bosons (such as bosonic nuclei) and fermions each have their
own power law exponent 
\begin{math}  \alpha  \end{math} in the high energy regime depending 
purely upon statistics. A likely source  of these evaporating cosmic 
rays are compact stellar objects such as neutron stars. Such objects 
would also radiate a copious amount of electrons (directly or from neutron decay)
and electron-positron pairs (if for
no other reason than that fast charged particles, be they electrons or baryons, 
when scattering through any background matter or radiation will produce 
such pairs). It is of central importance that the AMS collaboration 
has measured the appropriate fermion power law exponent 
here characteristic of the lepton sector of cosmic rays.

\section{Acknowledgements}

J. S. would like to thank the United States National Science Foundation for support under PHY-1205845.
Y. S. would like to thank Professor Bruna Bertucci for helpful discussions.

\end{document}